# DNA looping: the consequences and its control

Leonor Saiz and Jose M. G. Vilar[*]

Integrative Biological Modeling Laboratory, Computational Biology Program, Memorial Sloan-Kettering Cancer Center, 1275 York Avenue, Box #460, New York, NY 10021, USA

[*]Correspondence: vilar@cbio.mskcc.org

**Abstract**

The formation of DNA loops by proteins and protein complexes is ubiquitous to many fundamental cellular processes, including transcription, recombination, and replication. Here we review recent advances in understanding the properties of DNA looping in its natural context and how they propagate to the cellular behavior through gene regulation. The results of connecting the molecular properties with cellular physiology indicate that looping of DNA *in vivo* is much more complex and easier than predicted from current models and reveals a wealth of previously unappreciated details.

**Introduction**

DNA looping is deeply involved in many cellular processes, such as transcription, recombination, and replication [1-4], allowing distal DNA regions to affect each other. It is especially prominent in the regulation of gene expression, where proteins bound far away from the genes they control can be brought to the initiation of transcription region by looping the intervening DNA. The interplay between DNA looping and gene regulation was first identified in the *E. coli ara* operon [5], although it was already suspected to be present in eukaryotic enhancers [6] and in prokaryotic transcription [7]. Since then, it has been found in many other systems, such as the *gal*, *lac*, and *deo* operons in *E. Coli* [1,2], the lysogenic to lytic switch in phage λ [8], and the human β-goblin locus [9]. Recent examples show that it is present even in RXR [10] and p53 [11], two proteins widely involved in cancer.

Full understanding of the integration of DNA looping into such a diversity of cellular processes requires quantitative approaches. A key quantity is the free energy of looping DNA, which determines how easily DNA can loop and therefore the extent to which distal DNA sites can affect each other [4]. Through this quantity, DNA looping can easily be incorporated into thermodynamic models for the assembly of DNA-protein complexes that control different cellular processes. In this review we discuss recent advances in understanding the *in vivo* properties of DNA looping and their implications for gene regulation. We consider first the *in vivo* molecular properties of the looping process and examine their salient features, the differences with the *in vitro* data, and the expectations of current elastic DNA models. We then sketch briefly the key thermodynamic concepts needed to develop quantitative models for DNA-protein complexes and explore the consequences of DNA looping in gene regulation.

**Two types of loops**

DNA loops can be classified into two main categories with a fuzzy boundary: short or energetic (Figures 1a and 1b) and long or entropic (Figure 1c). This distinction comes from the physical forces that dominate their formation. For short loops, with lengths shorter than the DNA persistence length (~150 bp), the main determinant of looping is DNA elasticity. Thus, bending and twisting DNA, as well as the elastic properties of the molecules that tie the loop, play an important role. For long loops, in contrast, the limiting step is the erratic motion in the cell of the two DNA regions before they find each other. Thus, the main determinant is the lost of entropy that happens when two DNA regions are tied together.

Current theories [12-14] and most *in vitro* experiments [14-16] indicate that formation of short and long loops is extremely costly. And yet, short and long DNA loops are widely present *in vivo*. They can be as small as 60 bp in the *lac* operon [17], or 80 bp in nucleosome wrapping [18], and as long as 180 kb in mating type switching in yeast [19].

How does the intracellular environment mediate the formation of such loops? The first step to address this question is to obtain the properties of the cellular components in their natural environment. The extreme complexity of the cell, however, poses a strong barrier for experimentally characterizing the cellular components, not only because the properties of the components can change when studied *in vitro* outside the cell but also because the *in vivo* probing of the cell can perturb the process under study [20].

## From cellular physiology to *in vivo* molecular properties

A combined computational-experimental approach has recently been used to infer the *in vivo* free energies of DNA looping by the *lac* repressor [21] from measurements of enzyme production in the *lac* operon [17] for different lengths of the loop. The key idea is to use a well-established mathematical model for the regulation of gene expression in the *lac* operon "in reverse". In this way, it is possible to go from the observed cellular behavior to the properties of the unperturbed cellular components. The free energy of looping by the *lac* repressor for the specific experimental conditions analyzed with this approach [21] follows from the concise expression

$$\Delta G_l = -RT \ln \frac{R_{loop} - R_{noloop}}{R_{noloop} - 1}[N], \qquad (1)$$

where $R_{loop}$ is the measured repression level, a dimensionless quantity used to quantify the extent of repression of a gene; $R_{noloop}$ is the repression level in the absence of DNA looping; $[N]$ is the concentration of repressors; and $RT$ is the gas constant times the absolute temperature ($RT \approx 0.6$ kcal/mol for typical experimental conditions). The results obtained present marked differences with the current *in vitro* view of DNA looping.

For short loops (Figure 1d), this analysis showed that the free energy of looping oscillates with the helical periodicity of DNA (~10.9 bp) as the length of the loop changes, which was expected because the operators must have the right phase to bind simultaneously to the repressor, and, unexpectedly, that the free energy in a cycle behaves asymmetrically [21]. This asymmetry is characterized by a second representative oscillatory component with a period of ~5.6 bp. Other striking features are that the amplitude of the oscillations is extremely small, ~2.5 kcal/mol, and that the *in vivo* free energy does not seem to diverge for short loop lengths. These results indicate that the formation of *in vivo* DNA loops is much more complex and easier than expected from current theories, which predict symmetric and, at least, twice as big oscillations [14,22].

For long loops (Figure 1e), the resulting *in vivo* free energy of looping nicely fits the theoretically predicted expression for a flexible polymer $\Delta G_{l_0} + \alpha RT \ln(l/l_0)$, where $l$ is the length of the loop, $l_0$ is a reference length and $\alpha$ is a constant [4]. Intriguingly, theoretical estimates give $\alpha \approx 2.25$ [12,13], which is significantly different from the inferred in vivo value $\alpha \approx 1.24$ ($1.24 RT \ln(l) + 4.72$). This result is even more remarkable because the theoretical lower bound of this parameter for loop formation in 3 dimensions is $\alpha = 1.5$, the value for an ideal polymer without excluded volume effects. As in the case of short loops, here, the *in vivo* environment seems to facilitate also the formation of long DNA loops.

## *In vivo* intricacies

The origin of the differences between the predictions of continuum elastic models and the observed *in vivo* behavior remains far from being fully resolved. Recent structural and computational studies on DNA [18,23] indicate that the loop can be bent and twisted nonuniformly because of different contributions, such as, for instance, the anisotropic flexibility of DNA, local features resulting from the DNA sequence, and interactions with the *lac* repressor [24] and other DNA binding proteins [25]. The formation of DNA loops is also tightly coupled to the molecular properties of the proteins and protein complexes that form the loop. Moreover, depending on the orientation of the two DNA binding sites and the properties of the looped DNA-protein complex, the DNA loop can be accomplished following different trajectories [25-27].

Only very recently it has become clear that the *in vivo* behavior for short loops (Figure 1d) can be accurately accounted for by the simultaneous presence of two distinct conformations of the looped DNA-protein complex [28]. These two conformations have different bending and torsional properties. As the length of the loop changes, the less stable conformation becomes the most stable one. This alternating pattern is repeated periodically and different loop conformations are adopted in each case for DNA to find the configuration with the minimum free energy. It is also possible to use the formula for the free energy as a function of the repression level (Equation 1) with data for different mutants [29] to infer the effects of key architectural properties on DNA. When the HU protein, which helps bending of DNA, is absent in the cell, the free energy of looping DNA increases and the oscillations become symmetric [28]. In all cases studied, two wild type-like and one mutant strain, there are present the contributions of at least two conformations.

The properties obtained by fitting the inferred *in vivo* data [28] with a elastic model with two conformations are consistent with those obtained with a recent theory of sequence-dependent DNA elasticity for the lac repressor-DNA complex [30]. This computational approach and the inferred *in vivo* data together highlight the need for more detailed models of DNA looping. The inferred high versatility of looped DNA-protein complexes at establishing different conformations in the intracellular environment seems to underlie the unanticipated behavior of the *in vivo* free energy of DNA looping for short loop lengths and can be responsible not only for asymmetric oscillations with decreased amplitude but also for plateaus and secondary maxima (Fig. 1d). All these features indicate that the physical properties of DNA can actively be selected for controlling the cooperative binding of regulatory proteins and achieving different cellular behaviors.

## Two modes of looping

The study of the induction switches in phage λ and the *lac* operon led to the discovery of gene regulation [8,31]. As it turned out, both systems rely on DNA looping [32-34]. They exemplify two main modes of forming DNA loops. In the *lac* operon, DNA looping is mediated by the simultaneous binding of the two DNA binding domains of a single repressor molecule to two DNA sites known as operators [35]. In phage λ, in contrast, the loop is not formed by a single protein but by a protein complex that is assembled on DNA when the loop forms [33].

These two modes of looping are present in many systems. For instance, a pattern of induced cooperativity similar to that of phage λ is observed in RXR, a nuclear hormone receptor [10]. In its tetrameric form, RXR has two DNA binding domains and can loop DNA to bring transcription factors close to the promoter region. Retinoic acid controls whether or not the loop is formed by preventing the assembly of the tetrameric complex from the constituent dimers, which can also bind DNA. The E2 transactivator protein of bovine papilloma virus, on the other hand, loops DNA following the looping mode of the *lac* repressor [36]. Remarkably, if more than two binding sites are present on the same strand of DNA, E2 can even form multiple simultaneous loops that are visible by electron microscopy [36].

In general, multiple proteins are assembled to form functional complexes on looped DNA. In eukaryotic transcription, for instance, there are multiple DNA binding sites spread over long distances that are involved in controlling the same localized DNA events. DNA looping in this case allows multiple proteins to affect the RNA polymerase in the promoter region. Enhancers, silencers, or mediators bound at distal DNA sites are then brought to form part of, affect, or interfere with the transcriptional complex. Understanding this type of molecular complexity requires quantitative approaches that extend beyond prototypical chemical reactions in a well-stirred reactor [4].

**The quantitative approach**

DNA looping is typically controlled by the interaction of proteins with DNA to form dynamic nucleoprotein complexes. The most widely used quantitative approaches to study DNA-protein assembly are based on thermodynamics [37]. Thermodynamics allows for a straightforward connection of the molecular properties of the system with the effects that propagate up to the cellular physiology. Each configuration $s$ of the DNA-protein complex has associated a free energy $\Delta G(s)$, which is connected to the equilibrium probability $P_s$ of such configuration through the statistical interpretation of thermodynamics; namely, $P_s = \frac{1}{Z} e^{-\Delta G(s)/RT}$, where $Z = \sum_s e^{-\Delta G(s)/RT}$ is the normalization factor [37].

The key quantities to understand the control of DNA looping are positional, interaction, and conformational free energies [4]. The positional free energy, $p$, accounts for the cost of bringing one component to the protein-DNA complex, for instance bringing the *lac* repressor to its DNA binding site. Its dependence on the component concentration, $[N]$, is given by $p = p_0 - RT \ln[N]$, where $p_0$ is the positional free energy at 1M. This type of dependence indicates that it is easier to bring a component into the complex if its concentration is higher. Interaction free energies, $e$, arise from the physical contact between components (e.g., electrostatic interactions) and conformational free energies, $c$, account for changes in conformation (e.g., looped *vs.* unlooped states). Typical values (in kcal/mol) for the *in vivo* DNA-*lac* repressor complex are $p \approx 26$, $e \approx -28$, and $c \approx 23$. Two key points are that the different contributions can be positive or negative and that typically their absolute values are much larger than the thermal energy ($\approx 0.6$). By collecting all the contributions to the free energy, it is possible to infer the dominant conformation of the protein-DNA complex for each specific condition, which corresponds to the one with the smallest free energy.

To illustrate these concepts in more detail, we consider the binding of the bidentate *lac* repressor to two operators, *O*1 and *O*2 (Figure 2a). The *lac* repressor-DNA complex can be in five representative states [38]: (i) none of the operators is occupied, (ii) a repressor is bound to just *O2*, the auxiliary operator, (iii) a repressor is bound to just *O1*, the main operator, (iv) a repressor is bound to both *O1* and *O2* by looping the intervening DNA, and (v) two repressors are bound, one to each operator. The free energies for each of these states are $\Delta G_i = 0$, $\Delta G_{ii} = p + e_2$, $\Delta G_{iii} = p + e_1$, $\Delta G_{iv} = p + e_1 + e_2 + c_L$, and $\Delta G_v = 2p + e_1 + e_2$, respectively. Here, the quantity $p$ is the positional free energy of the repressor and embeds the dependence on the repressor concentration $[N]$; $e_1$ and $e_2$ are the interaction free energy between the repressor and *O1* and *O2*, respectively; and $c_L$ is the conformational free energy of looping DNA ($c_L \equiv p_0 + \Delta G_l$).

These free energies can be used to derive the probabilities of the different states (Figure 2b). For instance, the looped state (iv) is more probable than the one-repressor unlopped state (iii) if $e_2 < c_L$; that is to say, looping will be favored whenever establishing a second binding contact is less costly than looping DNA. In this case, DNA looping increases the occupancy of the DNA binding sites. If $p < c_L$, the looped state (iv) is more probable than the two-repressor unlopped state (v). This inequality is remarkable because it also indicates that the looped state is not favored for sufficiently high repressor concentrations. Thus, the repressor is responsible for forming the loop at low-moderate concentrations and for preventing it at high concentrations (Figure 2b).

Straightforward application of the standard thermodynamic approach [39] in a general framework is of limited use because the number of states that must be considered typically increases exponentially with the number of components. It has become clear recently that it is possible to overcome this limitation and express the free energy of all these states in a compact form by using binary variables [40]. In the case of the *lac* operon, this new approach leads to

$$\Delta G(s) = (p + e_1)s_1 + (p + e_2)s_2 + (c_L - ps_1s_2)s_L, \qquad (2)$$

where $s_1$ and $s_2$ are binary variables that indicate whether ($s_i = 1$; for $i = 1, 2$) or not ($s_i = 0$; for $i = 1, 2$) the repressor is bound to *O1* and *O2*, respectively; and $s_L$ is a variable that indicates the conformational state of the DNA, either looped ($s_L = 1$) or unlooped ($s_L = 0$). Thus, it is possible to write a global concise expression, instead of one for each of the five states, to specify the thermodynamic properties of the system. This expression can be used to compute different static and dynamic quantities without having to instantiate explicitly all the potential states [40].

## How fast?

The dynamic properties of DNA are also important in many processes, for instance, in controlling transcriptional noise [4]. The relationship between kinetic and thermodynamic properties known as the principle of detailed balance can be exploited to infer the rate of loop formation, $k_{loop}$ [38]. Assuming that the dissociation rate of one repressor domain from DNA does not depend on whether the other domain is bound to DNA, it leads to

$$k_{loop} = k_a e^{-\Delta G_l / RT}, \qquad (3)$$

where $k_a = 8.8 \times 10^7 M^{-1} s^{-1}$ is the association rate constant for the binding of the repressor to the operator, which for $\Delta G_l = 8.4$ kcal/mol results in $k_{loop} = 74 s^{-1}$ [38]. Thus, unlooped DNA with the repressor bound to one operator reloops within 10-20 ms. This time scale is similar to that for looping of DNA around nucleosomes, where unwrapped DNA rewraps within approximately 10-50 ms [41].

## The effects

DNA looping has many obvious effects because of its role in mediating long range interactions on DNA. It allows two, or more, DNA regions that are far apart to come close to each other, which is needed, for instance, to allow the transfer of genetic information that happens during recombination [19,42]. DNA loops are also used to tie the end of chromosomes and regulate the length of telomeres [43]. Beyond these systems in which it is strictly required, DNA looping is also used to increase the strength of binding of regulatory molecules to their cognate sites. The thermodynamic approach we have discussed shows how such increase is achieved in the *lac* operon, where the looped state is always more stable than both unlooped states with one repressor (Figure 2b). DNA looping has also other more subtle roles, which are strongly interrelated with the inherent stochastic nature of cellular processes.

Computational modeling of the *lac* operon [38] together with experimental data [44] strongly suggests that DNA looping can be used to decrease the sensitivity of transcription to changes in the number of regulatory proteins. The transcription rate in the *lac* operon for the looping case shows a plateau-like behavior, centered around 50 nM, which is not present in the regulation with just a single operator (Figure 2c). The low sensitivity obtained with DNA looping in this region can be used to achieve fairly constant transcription rates among cells in a population, irrespective of the fluctuations in the numbers of *lac* repressor molecules. In contrast, using a single operator just propagates the fluctuations proportionally.

DNA looping can also reduce the intrinsic fluctuations of transcription [38]. If transcription switches slowly between active and inactive, there are long periods of time in which proteins are produced constantly and long periods without any production. Therefore, the number of molecules fluctuates strongly between high and low values. In contrast, if the switching is very fast, the production happens in the form of short and frequent bursts. This lack of long periods of time with either full or null production gives a narrower distribution of the number of molecules. DNA looping naturally introduces a fast time scale: the time for the repressor to be recaptured by the main operator before unbinding the auxiliary operator, which, as we have shown above, is much shorter than the time needed by a repressor in solution. Therefore, DNA properties are also important for controlling transcriptional noise.

**Conclusions**

DNA looping is an extremely important process for the functioning of even the simplest types of cells. Besides providing a backbone for fundamental long range interactions,

DNA looping can be used to increase specificity and affinity simultaneously, and, at the same time, to control the intrinsic stochasticity of cellular processes. In particular, it can buffer molecular variability to produce phenotypically homogeneous populations and decrease the transcriptional noise [4].

It is becoming increasingly clear that the cell has found ways to loop DNA that extend beyond the classical view of an extremely stiff polymer at short length scales. Recent approaches connecting cellular physiology measurements with the *in vivo* free energy of looping DNA by the *lac* repressor indicate that DNA loops can form extremely easily in the intracellular environment: the *in vivo* free energy of looping DNA changes within a very narrow window of about 2.5 kcal/mol over loop lengths that range from 50 bp to 1.5 kb (Figures 1d and 1e). These changes in the free energy are much smaller than predicted from current DNA elastic models and lie between the typical values of the free energies of interaction between regulatory molecules [45].

The properties of *in vivo* looping DNA seem to have been tuned for the effects of regulatory molecules to be strongly dependent on their precise DNA positioning and at the same time easily tunable and modifiable by their cooperative interactions. At the intracellular level, the looping properties of DNA are affected, among other factors, by the sequence dependence of DNA elasticity, presence of alternative loop conformations, interactions with different proteins, and DNA supercoiling [14]. Understanding how all these factors are combined to obtain the observed behavior is one of the main challenges that lies ahead.

**References**

1. Adhya S: **Multipartite genetic control elements: communication by DNA loop**. *Annu Rev Genet* 1989, **23**:227-250.
2. Schleif R: **DNA looping**. *Annu Rev Biochem* 1992, **61**:199-223.
3. Matthews KS: **DNA looping**. *Microbiol Rev* 1992, **56**:123-136.

4. Vilar JMG, Saiz L: **DNA looping in gene regulation: from the assembly of macromolecular complexes to the control of transcriptional noise**. *Current Opinion in Genetics & Development* 2005, **15**:136-144.
5. Dunn TM, Hahn S, Ogden S, Schleif RF: **An operator at -280 base pairs that is required for repression of araBAD operon promoter: addition of DNA helical turns between the operator and promoter cyclically hinders repression**. *Proc Natl Acad Sci U S A* 1984, **81**:5017-5020.
6. Moreau P, Hen R, Wasylyk B, Everett R, Gaub MP, Chambon P: **The SV40 72 base repair repeat has a striking effect on gene expression both in SV40 and other chimeric recombinants**. *Nucleic Acids Res* 1981, **9**:6047-6068.
7. Irani MH, Orosz L, Adhya S: **A control element within a structural gene: the gal operon of Escherichia coli**. *Cell* 1983, **32**:783-788.
8. Ptashne M: *A genetic switch: phage lambda revisited* edn 3rd. Cold Spring Harbor, N.Y.: Cold Spring Harbor Laboratory Press; 2004.
9. Tolhuis B, Palstra RJ, Splinter E, Grosveld F, de Laat W: **Looping and interaction between hypersensitive sites in the active beta-globin locus**. *Mol Cell* 2002, **10**:1453-1465.
10. Yasmin R, Yeung KT, Chung RH, Gaczynska ME, Osmulski PA, Noy N: **DNA-looping by RXR tetramers permits transcriptional regulation "at a distance"**. *J Mol Biol* 2004, **343**:327-338.
11. Stenger JE, Tegtmeyer P, Mayr GA, Reed M, Wang Y, Wang P, Hough PVC, Mastrangelo IA: **P53 Oligomerization and DNA Looping Are Linked with Transcriptional Activation**. *Embo Journal* 1994, **13**:6011-6020.
12. Rippe K: **Making contacts on a nucleic acid polymer**. *Trends Biochem Sci* 2001, **26**:733-740.
13. Hanke A, Metzler R: **Entropy loss in long-distance DNA looping**. *Biophys J* 2003, **85**:167-173.
14. Bloomfield VA, Crothers DM, Tinoco I: *Nucleic Acids: Structures, Properties, and Functions*. Sausalito, CA: University Science Books; 2000.
15. Du Q, Smith C, Shiffeldrim N, Vologodskaia M, Vologodskii A: **Cyclization of short DNA fragments and bending fluctuations of the double helix**. *Proc Natl Acad Sci U S A* 2005, **102**:5397-5402.
16. Cloutier TE, Widom J: **Spontaneous sharp bending of double-stranded DNA**. *Mol Cell* 2004, **14**:355-362.
17. Muller J, Oehler S, Muller-Hill B: **Repression of lac promoter as a function of distance, phase and quality of an auxiliary lac operator**. *J Mol Biol* 1996, **257**:21-29.
18. Richmond TJ, Davey CA: **The structure of DNA in the nucleosome core**. *Nature* 2003, **423**:145-150.
19. Broach JR: **Making the right choice--long-range chromosomal interactions in development**. *Cell* 2004, **119**:583-586.
20. Alberts B: *Molecular biology of the cell* edn 4th. New York: Garland Science; 2002.
21. Saiz L, Rubi JM, Vilar JMG: **Inferring the in vivo looping properties of DNA**. *Proceedings of the National Academy of Sciences of the United States of America* 2005, **102**:17642-17645.

22. Yan J, Marko JF: **Localized single-stranded bubble mechanism for cyclization of short double helix DNA**. *Phys Rev Lett* 2004, **93**:108108.
23. Olson WK, Swigon D, Coleman BD: **Implications of the dependence of the elastic properties of DNA on nucleotide sequence**. *Philos Transact A Math Phys Eng Sci* 2004, **362**:1403-1422.
24. Villa E, Balaeff A, Schulten K: **Structural dynamics of the lac repressor-DNA complex revealed by a multiscale simulation**. *Proc Natl Acad Sci U S A* 2005, **102**:6783-6788.
25. Semsey S, Virnik K, Adhya S: **A gamut of loops: meandering DNA**. *Trends Biochem Sci* 2005, **30**:334-341.
26. Halford SE, Gowers DM, Sessions RB: **Two are better than one**. *Nature Structural Biology* 2000, **7**:705-707.
27. Zhang Y, McEwen AE, Crothers DM, Levene SD: **Statistical-mechanical theory of DNA looping**. *Biophys J* 2006, **90**:1903-1912.
28. Saiz L, Vilar JMG: **In vivo evidence of alternative loop geometries in DNA-protein complexes**. *arXiv:q-bio.BM/0602012* 2006.
29. Becker NA, Kahn JD, Maher LJ, 3rd: **Bacterial repression loops require enhanced DNA flexibility**. *J Mol Biol* 2005, **349**:716-730.
30. Swingon D, Coleman BD, Olson WK: **upublished**. *preprint* 2006.
31. Müller-Hill B: *The lac Operon: a short history of a genetic paradigm.* Berlin; New York: Walter de Gruyter; 1996.
32. Mossing MC, Record MT, Jr.: **Upstream operators enhance repression of the lac promoter**. *Science* 1986, **233**:889-892.
33. Revet B, von Wilcken-Bergmann B, Bessert H, Barker A, Muller-Hill B: **Four dimers of lambda repressor bound to two suitably spaced pairs of lambda operators form octamers and DNA loops over large distances**. *Curr Biol* 1999, **9**:151-154.
34. Dodd IB, Shearwin KE, Perkins AJ, Burr T, Hochschild A, Egan JB: **Cooperativity in long-range gene regulation by the lambda CI repressor**. *Genes Dev* 2004, **18**:344-354.
35. Lewis M, Chang G, Horton NC, Kercher MA, Pace HC, Schumacher MA, Brennan RG, Lu P: **Crystal structure of the lactose operon repressor and its complexes with DNA and inducer**. *Science* 1996, **271**:1247-1254.
36. Knight JD, Li R, Botchan M: **The activation domain of the bovine papillomavirus E2 protein mediates association of DNA-bound dimers to form DNA loops**. *Proc Natl Acad Sci U S A* 1991, **88**:3204-3208.
37. Hill TL: *An introduction to statistical thermodynamics.* Reading, Mass.: Addison-Wesley Pub. Co.; 1960.
38. Vilar JMG, Leibler S: **DNA looping and physical constraints on transcription regulation**. *J Mol Biol* 2003, **331**:981-989.
39. Ackers GK, Johnson AD, Shea MA: **Quantitative Model for Gene-Regulation by Lambda-Phage Repressor**. *Proceedings of the National Academy of Sciences of the United States of America-Biological Sciences* 1982, **79**:1129-1133.
40. Saiz L, Vilar JMG: **Stochastic dynamics of macromolecular-assembly networks**. *Molecular Systems Biology* 2006, **in press**.

41. Li G, Levitus M, Bustamante C, Widom J: **Rapid spontaneous accessibility of nucleosomal DNA**. *Nat Struct Mol Biol* 2005, **12**:46-53.
42. Radman-Livaja M, Biswas T, Ellenberger T, Landy A, Aihara H: **DNA arms do the legwork to ensure the directionality of lambda site-specific recombination**. *Curr Opin Struct Biol* 2006, **16**:42-50.
43. de Lange T: **T-loops and the origin of telomeres**. *Nat Rev Mol Cell Biol* 2004, **5**:323-329.
44. Oehler S, Amouyal M, Kolkhof P, von Wilcken-Bergmann B, Muller-Hill B: **Quality and position of the three lac operators of E. coli define efficiency of repression**. *Embo J* 1994, **13**:3348-3355.
45. Ptashne M, Gann A: *Genes & signals*. Cold Spring Harbor, N.Y.: Cold Spring Harbor Laboratory Press; 2002.

**Annotations to references**

Reference [4]**

This reference puts forward the thermodynamic concepts underlying macromolecular assembly, with emphasis on the formation of protein-DNA complexes with loops.

Reference [17]*

The authors systematically varied the distance between two operators in the *lac* operon (from 57.5 to 98.5 bp in increments of 1 bp and from 100 to 1500 bp for representative values) and measured the repression levels under conditions similar to wild type. The measured repression levels can be used to compute the free energy of looping DNA *in vivo*.

Reference [21]**

The authors infer the *in vivo* free energy of looping DNA by the *lac* repressor for different lengths of the loop. Strikingly, in addition to the intrinsic periodicity of the DNA double helix, the *in vivo* free energy has an oscillatory component of about half the helical period. The total amplitude of the oscillations is also much smaller than predicted from current models.

Reference [28]**

The authors develop a concise model that incorporates two elastic conformations of the *lac* repressor-DNA complex. This model accounts in full detail for the *in vivo* behavior of the free energy for short loops.

Reference [29]*

This reference follows a similar approach to that of Muller et al. [17] but considers also mutants that lack key architectural proteins. From the reported data, it was possible to infer in Ref. [28] the effects of HU proteins on the looping properties of DNA.

Reference [30]**

The authors use a recent theory of sequence-dependent DNA elasticity to compute the free energy of looping of the *lac* repressor-DNA complex for different conformations of the complex and different lengths of the loop. The results show an excellent agreement with the results of the analysis of the *in vivo* data of Ref. [28], including the amplitude of the oscillations and the lack of short-loop divergence of the free energy.

Reference [40]**

The authors integrate the thermodynamic concepts of Ref. [4] in a mathematical approach for computing the stochastic dynamics of macromolecular assembly, which includes DNA loops formed by protein and protein complexes and their effects in gene regulation.

Reference [41]**

The authors measure the rates of wrapping and unwrapping nucleosomal DNA. The results indicate that they are very fast, which explains how remodeling factors can be recruited to particular nucleosomes on a biologically relevant timescale.

**Figure Legends**

FIGURE 1. Looped conformations and the *in vivo* free energy of looping DNA by the *lac* repressor. The bidentate *lac* repressor (shown in red) can loop DNA (orange thick line) in different ways: **(a)** short loop with repressor in a V-shape conformation; **(b)** short loop with repressor in an extended conformation; and **(c)** long loop with supercoiled DNA. The *in vivo* free energy of looping DNA [21] as a function of the length of the loop for **(d)** short and **(e)** long loops has been obtained using a computational-experimental approach (red square symbols) as described in Saiz et al. [21] (see text) from the measured repression levels of Muller et al. [17]. For short loops, the thick black line represents the best fit to the looping free energy $\Delta G_l$ given by an elastic DNA model that considers the contributions of two loop conformations (Equations 1 and 2 of Ref. [28]). The two alternative loop conformations of the *lac* repressor-DNA complex could include two conformations of the *lac* repressor or two different binding motifs as represented in the cartoons. For long loops, the thick line represents the best fit using the theoretically predicted expression for an ideal flexible polymer: $1.24 RT \ln(l) + 4.72$, where $l$ is the length of the loop [4].

FIGURE 2. Relevant states for *lac* repressor binding to two operators, their probabilities, and their effect in transcription regulation. **(a)** The *lac* repressor binding to two operators has five representative states. The promoter (arrow), downstream the main operator, is repressed when the *lac* repressor (shown in red) is bound to the main operator (states (iii), (iv), and (v)) and unrepressed when the main operator is unoccupied (states (i) and (ii)). Binding to the auxiliary operator does not affect transcription. The thick black line represents DNA with the two *lac* operators shown as orange boxes. Here, $p$ is the positional free energy of the repressor, $e_1$ and $e_2$ are the interaction free energy between the repressor and the main and auxiliary operator, respectively; and $c_L \equiv p_0 + \Delta G_l$ is the conformational free energy of looping DNA. **(b)** The probability of the different states as a function of the repressor molar concentration $[N]$ has been obtained with the statistical thermodynamic approach, as described in the text. The values used for the different contributions to the free energy (in kcal/mol) are $e_1 = -28.1$, $e_2 = -26.6$, $p = 15 - 0.6\ln[N]$, and $c_L = 23.35$. Only the states with relevant populations are labeled. The looped state (iv) is the most abundant except for low and high repressor concentrations. **(c)** The normalized transcription rate as a function of the *lac* repressor concentration for one (blue circles and black dashed lines) and two (red squares and continuous black lines) operators shows an excellent agreement with the available experimental data [44]. The computed values of the normalized transcription rate $\tau = \frac{1}{Z}\sum_s (1-s_1)e^{-\Delta G(s)/RT}$ (lines) are compared with the experimental data (symbols) from Ref. [44] at two repressor concentrations for three different strengths of the main operator.

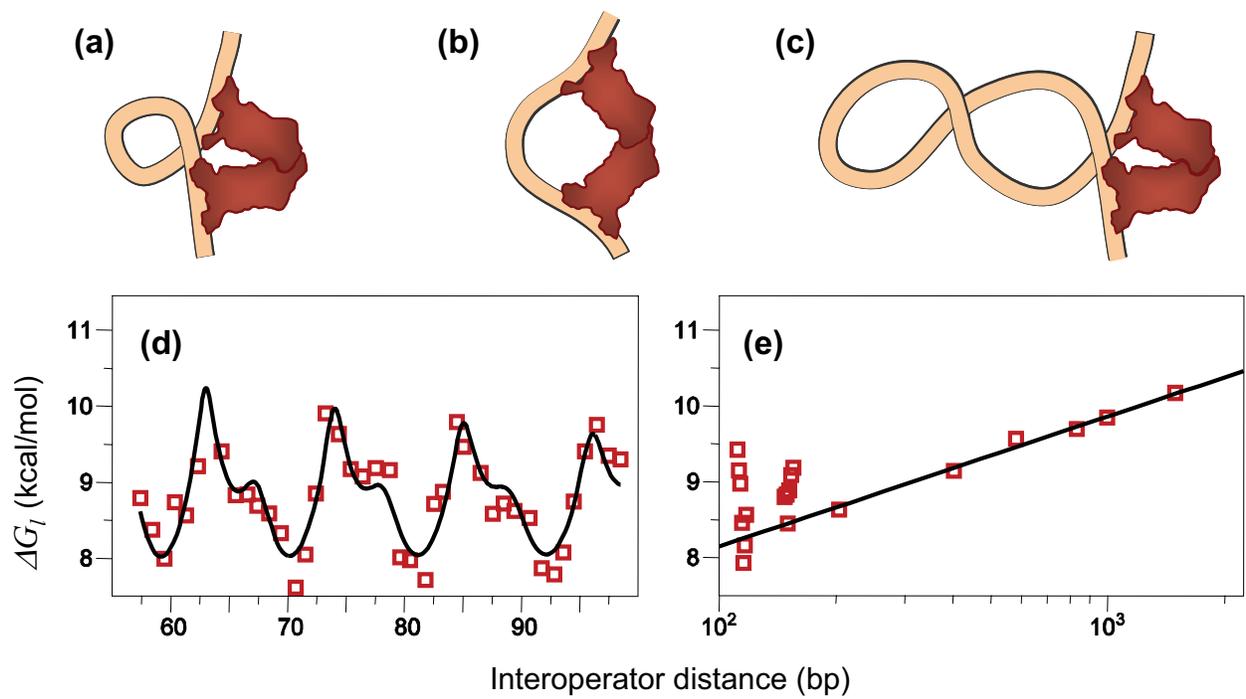

Figure 1

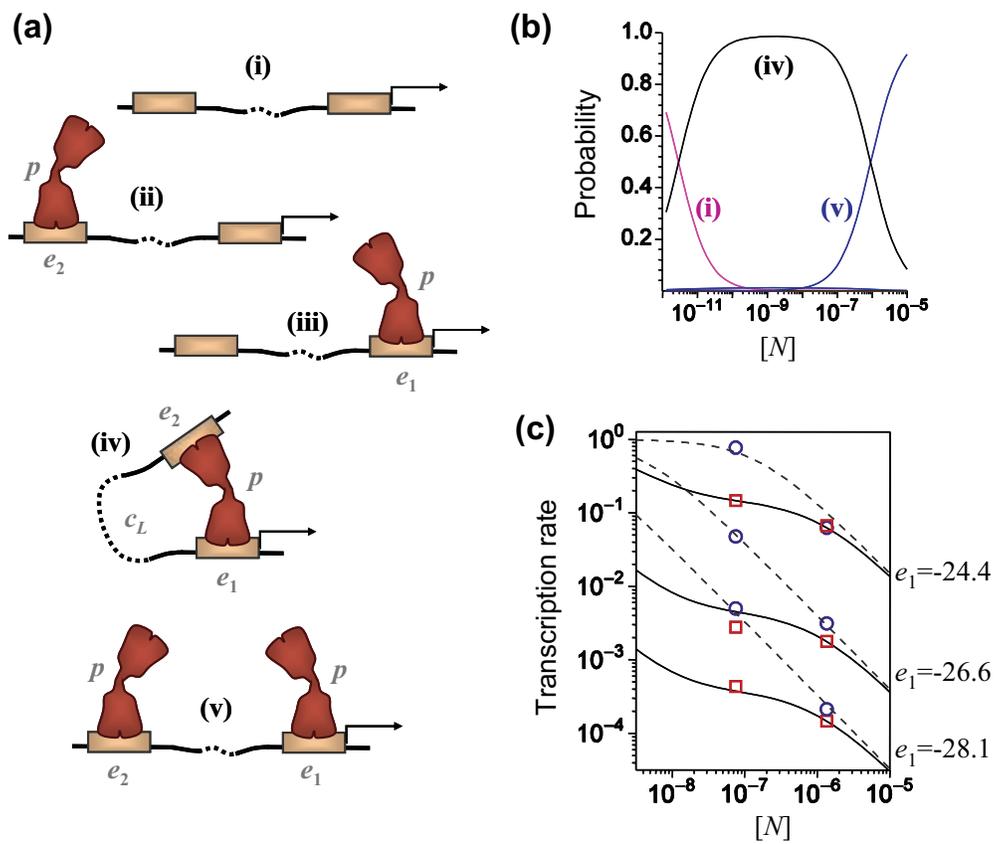

Figure 2